\documentclass[a4paper,UKenglish]{lipics}

\usepackage{microtype}

\newcommand{\bo}{\textbf}
\newcommand{\ep}{\emph}

\renewcommand{\t}{\textsc}

\newcommand{\noi}{\noindent}
\DeclareUnicodeCharacter{00A0}{~}

\title{ARG: A Virtual Tool for Teaching Argumentation Theory}
\titlerunning{ARG: A Virtual Tool for Teaching Argumentation Theory} 

\author[1]{Nailton Silva}
\author[2]{José Moura}
\author[3]{Patrick Terrematte}
\affil[1,2]{Academic Center of Rio Grande do Norte State (UNI-RN)\\  Law Department\\
 Natal -- RN -- Brazil\\
  \texttt{nailtongomes@ig.com.br}, \texttt{joseeduardomoura@unirn.edu.br}}
\affil[3]{Federal University of Rio Grande do Norte (UFRN)\\  Metropole Digital Institute (IMD)\\
Group of Logic, Language, Information, Theory and Applications (LoLITA) \\  Natal -- RN -- Brazil\\
  \texttt{patrickt@imd.ufrn.br}}
\authorrunning{N. Silva, J. Moura and P. Terrematte} 

\Copyright{Nailton Silva, José Moura and Patrick Terrematte}


\serieslogo{logo_ttl}
\volumeinfo
  {M. Antonia {Huertas}, Jo\~ao {Marcos}, Mar\'ia {Manzano}, Sophie {Pinchinat}, \\
  Fran\c{c}ois {Schwarzentruber}}
  {5}
  {4th International Conference on Tools for Teaching Logic}
  {1}
  {1}
  {207}
\EventShortName{TTL2015}

\begin{document}

\maketitle

\begin{abstract}
Researchers look for new virtual instruments that can improve and maximize traditional forms of teaching and learning. In this paper, we present the ARG system, a virtual tool developed to help the teaching/learning process in argumentation theory, especially in the field of Law. ARG was developed based on Araucaria by Reed and Rowe, Room 5 by Ronald P. Loui, as well on systems such as Argue!-System and ArguMed by Bart Verheij. ARG is a platform for online collaboration and applies the theory of Stephen Toulmin to produce arguments that are more concise, precise, minimally structured and more resistant to criticism.
\end{abstract}

\section{Introduction}

\begin{flushright}
\begin{minipage}{.6\textwidth}
\scriptsize
\emph{``The use of argumentation implies that one has renounced resorting to force alone, that value is attached to gaining the adherence of one's interlocutor by means of reasoned persuasion, and that one is not regarding him as an object, but appealing to his free judgment. Recourse to argumentation assumes the establishment of a community of minds, which, while it lasts, excludes the use of violence.'' }\\
\t{--- Chaim Perelman}
\end{minipage}
\end{flushright} 

\noi This study was initiated in mid-2010 to help undergraduate students understand the argumentative structures used by magistrates in the justification of their judicial decisions, adopting the accurate analysis of judgments pronounced by judges of the Rio Grande do Norte State Court as their methodology.
 
Following the senior thesis presented by Nailton Silva \cite{SILVA2014}, this study ccome to three findings, which were crucial to the development of ARG. The first being (i) identification of fragility in the composition of those Magistrates' arguments which were uncritical and did not observe rules pertinent to any argumentation theory.

Further, it was also observed that the substantiation of judicial decisions were being organized based on the same patterns that were already used in the very beginning of juridical rhetoric which accentuates, as an argumentative structure, the ``Legal Syllogism'',  creating the illusion of certainty in a sphere of uncertainty. Finally, it was noted that (iii) some arguments make use of assumptions and occult inference rules, establishing conclusions by implication. There are an obstacle for complete exercising of jurisdiction, of constitutional warrantees and credibility of the decisions and judicial institutions. Such conclusions served to be the justification for the search of higher rigor and consistency in the judicial determinations, and thus led to the development of the tool presented here.

We have analyzed how students of law are taught critical thinking and argumentation, focusing on argumentative structures used in the Brazilian legal practice. This analysis has led to the development of an electronic/virtual tool to assist in the task of argument construction.  The aim of this work is to present the ARG system\footnote{The ARG is available at \url{http://assistentearg.herokuapp.com/} with a video presentation at \url{http://youtu.be/BB0YDiOPam8}.}, a system designed to be a multilingual online tool able to assist in the production and improvement of  practical arguments -- not only those juridical in nature.

Our solution adopts Stephen Toulmin's theory of argumentation \cite{Toulmin2003,Toulmin2003a}, and we consider the value of his ``layout'' of argument, which enables us to visualize and follow the moves necessary for the construction of an argument of juridical practice. Toulmin's theory of argumentation, it is important to say, was already applied to compound electronic tools of representation and to helping in the composition of arguments\footnote{See \url{http://www.phil.cmu.edu/projects/argument_mapping/}.}, fitting very well when applied in numerous experiences of Artificial Intelligence and Argumentation and Law. This may be due to its heuristic properties, which is exploitable in information systems.

\section{State-of-the-art argumentation tools}

\begin{flushright}
\begin{minipage}{.6\textwidth}
\scriptsize
\emph{``We can’t understand where we are now, without understanding how we got there. And of course, once we recognize the need to understand how we get to a particular point, we have to recognize also that the work of our successors will supersede our own ideas, and we must be modest in recognizing that the best we can do is indeed the best we can do.'' }\\
\t{--- Stephen Toulmin. Reasoning in Theory and Practice, 2006}
\end{minipage}
\end{flushright} 

\noi There are well-known uses of informatics in Law\footnote{About the subject see: \url{http://austhink.com/critical/pages/argument_mapping.html} and \url{http://www.phil.cmu.edu/projects/argument_mapping/}.}, such as in the Mechanical Jurisprudence, CCLIPS - Civil Code Legal Information Processing System, JUDITH, British Nationality Act, HYPO, Zeno, Room 5, PROSUPPORT,  Araucaria and ArguMed projects. All of them essentially focus on the comprehension, representations and/or reproduction of judicial arguments. However, none of these serves as a solution for the problem highlighted in the previous section.

Four of these tools were the main tool available and have some of the functions of ARG. The first one was Room 5, developed by Ronald P. Loui, Jeff Norman, Joe Altepeter, Dan Pinkard, Dan Craven, Jessica Linsday and Mark Foltz, from Washington University. It was a website that provided a mechanism to represent, in a structured way, juridical argumentation, allowing visitors to make moves like those made by cases decided on the Supreme Court of the United States of America, inserted into a format that structures the discussion as either pro-author or pro-defendant \cite{Loui1997}.

The second and third systems that inspired us were ARGUE!-SYSTEM and ArguMed\footnote{Available at: \url{http://www.ai.rug.nl/~verheij/aaa/}.}, which were systems developed by Bart Verheij \cite{Verheij1998,Verheij1999,Verheij2005}, from the University of Maastricht in order to mediate the process and the argumentation of one or more users. In ARGUE!, the user provides argument data, such as assumptions, questions and reasons, and the system determines the status of the justification: whether they were justified, unjustified, or neither \cite{Verheij1998}. Its successor, ArguMed, with a more transparent subjacent argumentation theory and a more intuitive interface, provides the user with the gradual construction of arguments, through the filling-in of patterns, each of them corresponding to an ``argument move'', such as adducing a reason or making a declaration \cite{Verheij1998}.

Lastly, Araucaria\footnote{Available at: \url{http://araucaria.computing.dundee.ac.uk/doku.php}.}, developed in 2001 by Chris Reed and Glenn Rowe \cite{Reed2004} from the University of Dundee, is a tool that helps in (re)construction, typesetting and evaluation of arguments through a graphical interface (``point-and-click'').

These tools, however, are limited to English speakers. Such applications do not include an interactive guide centered in on gradual composition and evaluation of arguments with the criterion of evaluation and improvement. They also do not provide an environment of discussion where students are able to explore the collective knowledge with technical support in theories of argumentation.

With analysis of the systems of automation, mediation and representation of reasoning encountered, it is possible to formulate the following conclusions that also serve the as basis of the new system: (i) arguments can be represented graphically through systems of electronic data processing; (ii) representation of arguments make their dialectical structure more evident; and (iii) theories of argumentation can be used in addition to an electronic tool to construct arguments.

In all argumentation situations we can use a methodology to clarify the structure of the arguments at stake. These tools previously mentioned took into account or created a theory of argumentation subjacent to their own judicial system. For similar purposes, ARG (abbreviation of ``\ep{argumentum}'') was developed: with the clear objective of helping in the composition and management of arguments, and to address the root of the problem of inconsistency and fragility in argumentation that occur in Brazilian judicial practice. As a secondary objective, our tool also intends on supporting the teaching of critical thinking to students of philosophy, literature and science, helping their comprehension of theories of argumentation. 

\section{The Argumentation Assistant: ARG}

\begin{flushright}
\begin{minipage}{.6\textwidth}
\scriptsize
\emph{``Arguments are human interactions through which such trains of reasoning are formulated, debated, and/or thrashed out.'' }\\
\t{--- Stephen Toulmin. An Introduction to Reasoning, 1979.}
\end{minipage}
\end{flushright} 

\noi The aim was to construct and make public an online a multilingual teaching tool on the theory of argumentation based on participatory methodology with open source-code that could make it possible to store arguments and to visualize them in a clear and concise way. We planned on making available interactive guides to help users learn the techniques of arguing well by providing them with problems that stimulate them to think and produce good arguments. For this, we used the following technologies: (i) Ruby 1.9.3p125; (ii) Ruby on Rails 3.2.3; (iii) PostgreSQL 9.1.3; (iv) Github; and (v) Heroku. These tools were suitable for the development of multilingual information systems, are integrated with social media, and also have a large repository of documentation available online.

\begin{figure}[htpb]
\centering
\caption{ARG's homepage.}
\includegraphics[width=14cm]{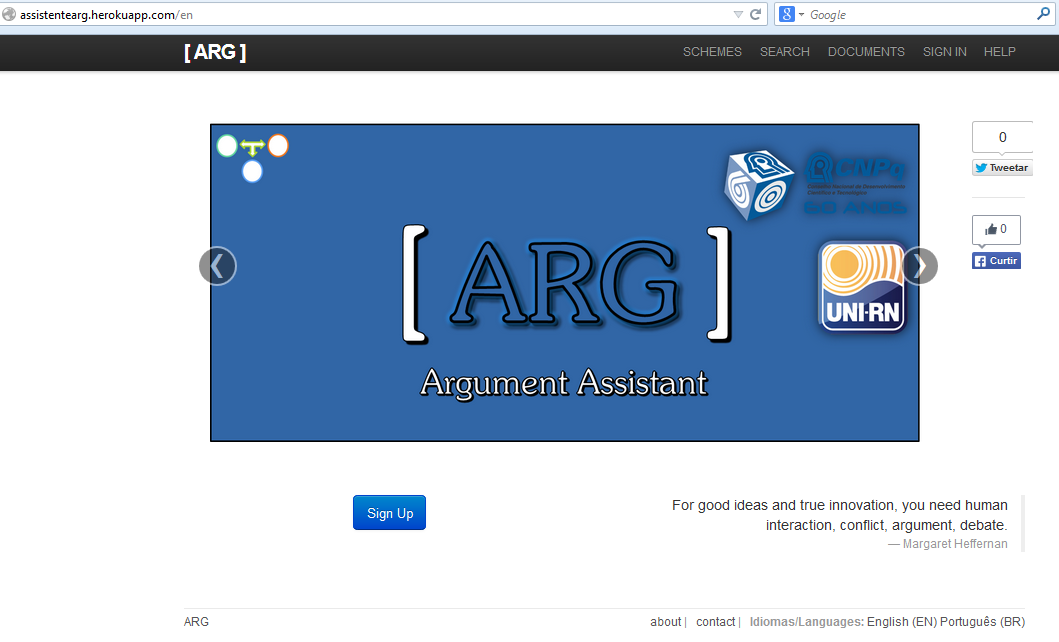}
\label{fig:arg-home}
\end{figure}

ARG is able to implement many functions. Some of them will be laid out briefly in this work, particularly the functions of (i) stimulation, (ii) evaluation, (iii) help/instruction and (iv) improving arguments.

The first function, stimulation, consists of prompting argument through cases provided by managers and/or professor-moderators of the system or by other academic students. Every week, for example, a new case might be introduced or a new problem regarding a previous case, opening a discussion where the users and moderators evaluate the arguments used, assigning a grade that it deems fitting, taking into account suggested criteria.

In ARG it is possible to post the abstract of a factual situation, attach elements of a police investigation and/or show a video and request, for example, that the users elaborate arguments that they have to compose. This could be an accusatory exordial (complaint) presented by the Prosecutor, for instance. After this, users could be asked to formulate the main argument of the answer to the accusation. 

All the arguments posted are stored in an external data bank and accessible in real time, with daily indexation. We have also created a repository of arguments oriented to cases and specific themes.

Regarding the second function, in the evaluation system, each user or moderator can, besides comment, attribute a maximum grade of five stars or a minimum of one star to any argument according to the quality of reasoning presented, ideally observing the presence of the following properties:

\begin{quote}
1. Clarity about what is  considered in the argument; 2. Necessary and sufficient data to support the conclusion; 3. Relevant support and proof of the case under discussion; 4. Strength of the conclusion explicitly gained and the possible refutations.  \cite{Toulmin2003a,Hitchcock2006} 

\end{quote}

In order to evaluate the arguments, it is necessary to observe that the following are avoided: ``1. Absent data; 2. Irrelevant data; 3. Deficient data; 4. Unjustified suppositions; 5. Ambiguity''\cite{Toulmin2003a}.

\begin{figure}[htpb]
\centering
\caption{ARG's page to guide the argumentative construction.}
\includegraphics[width=14cm]{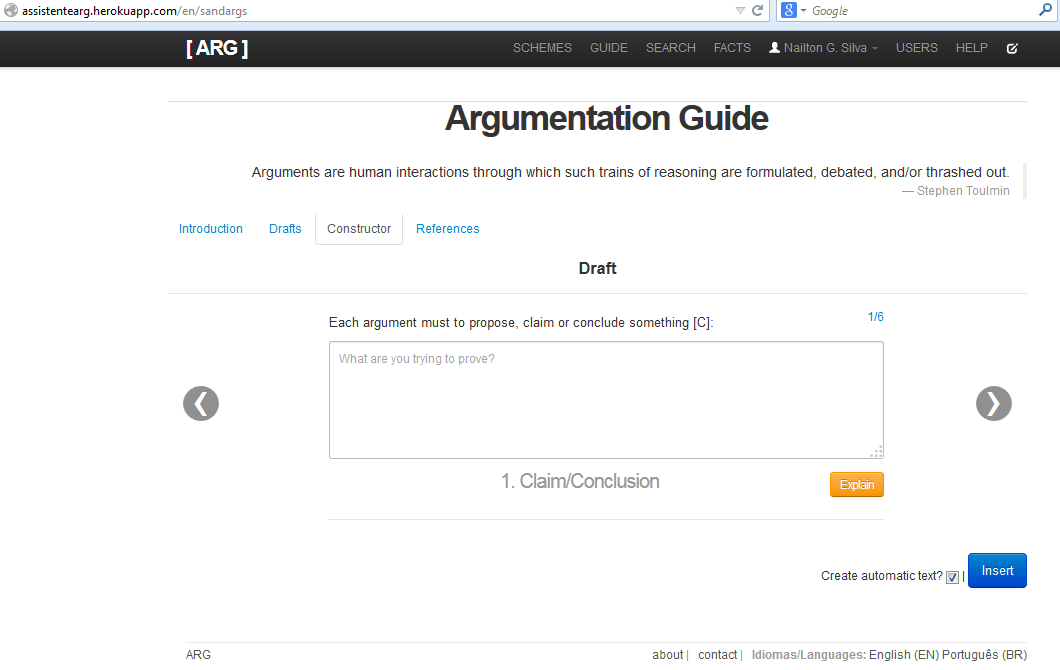}
\label{fig:arg-guide}
\end{figure}

In regards to the third function, in ARG it is possible to teach users the basics of any theory of argumentation, as it offers an interactive guide oriented by the theory of Toulmin (see Fig. \ref{fig:arg-guide}). Through this guide, it is requested that the user fills in the fields to generate a “skeleton” of an argument, treated by ARG as a ``draft'' or ``sketch''.

Finally, the fourth function focuses on improving arguments. ARG intends to help the prevention of eventual mistakes or neglect during its elaboration, disposing of a group of schemes of argumentation with many critical questions \cite{Walton2008}. Usually argument schemes have a number of critical questions that must be answered satisfactorily when there is its application in a particular case. Some of those questions refer to the acceptability of the premises and other issues related to the exceptional circumstances in which the scheme cannot be applied \cite{Prakken2005}.

\section{Solving a case with ARG: the Speluncean Explorers}

\begin{flushright}
\begin{minipage}{.6\textwidth}
\scriptsize
\emph{``The true revolution will arrive when computer systems now in development begin to perform or assist in the process of legal reasoning.'' }\\
\t{--- Computers and Legal Reasoning. Garry S. Grossman and Lewis D. Solomon, 1983.}\\
\end{minipage}
\end{flushright}

\noi As a hypothetical scenario, let  us suppose that we introduced on ARG the case of the Speluncean Explorers\footnote{The work describes the judgment of an appellation  resource where the accused were processed and condemned to death by the Court of Jury that recognized the practice of homicide. In synthesis, the victim was sacrificed by the accused to enable their survival, since they had been without any resources or food, involuntary and unpredictably for more than twenty days because of a landslide at the only access to a cave \cite{Fuller1949}.}, and that the character J. Keen, Magistrate of the Supreme Court, needs help in constructing an argument for his vote. 

There is a page in the system that provides a guide for argumentation and an argument ``constructor/builder''. In this case, the user is asked to fill in the appropriate fields with elements that, according to Toulmin, compound any argument. So, Keen, stimulated by questions\footnote{I.e. ``What are you trying to prove?'' or ``Why do you have those assumptions?''} and solicitations programmed in ARG, will fill the elements in with his arguments in the following way:

\begin{itemize}
  \item \bo{Claim} [C] = The sentence must be confirmed, with the maintenance of condemnation of the accused.
 \item \bo{Grounds}  [G] =  (1) It is not of competence for the judge to measure if the act was just or unjust, good or bad, but to apply the law and not their convictions of morality; 
      (2) The judiciary cannot create exceptions to the application of the law;
      (3) The judge cannot search the purposes of the law. In doing so, he legislates and, consequently usurps the competence of the Legislative Power; And
      (4) the conduct perpetrated by the accused fits the legal dispositive perfectly 
      \item \bo{Warrant} [W] = The judge must be loyal to the written law and interpret its evident meaning, without references to personal desire and its own concepts of justice.
\item \bo{Backing} [B] = Article 12-A of the criminal code.
\end{itemize}

The tool, considering the quantity of provided elements, will be able to generate an automatic text for the argument following the aforementioned guidelines:

\begin{quote}
This is based on article 12-A [B] of the criminal code. We assume that [W] the judge is be loyal to the written law and interprets it on its evident meaning, without interferences or personal desires and own conceptions of justice. Given [G]: (1) it is not the duty of the judge to measure if the act was just or unjust, good or bad, but to apply the law and not its conceptions of morality; (2) the judiciary cannot create exceptions to the applications of the law; (3) it is not up to the judge to search the purpose of the law, but under the penalty of usurping the competence of the Legislative Power; and (4) the conduct perpetrated by the accused fits the legal dispositive, therefore, [C] it must confirm the sentence, keeping the condemnation of the accused ones. 
\end{quote}

Subsequently, offers the user a roll of verification \footnote{With questions such as: 1. Is it clear and sure the claim the argument tries to make? 2. Is there clarity and certainty in what is implicitly purposed (is there any purpose)?; 3. Are the data/reasons relevant?; 4. Are the data/reasons enough?; 5. Are the data/reasons justified?; among others.} so that it is possible for the improvement of the argument. This roll corresponds to the structured abstract of what is necessary to compose a good argument, based on Toulmin \cite{Toulmin2003a} and Hitchcock \cite{Hitchcock2006}.

Thus, it is up to the user to verify the nature of the argument that it intends to propose and to observe if it is proper to the form and to the main proper questions. So, J. Keen, the judge in this example, could consult the scheme ``Argument from the Constitution of Causal Laws''\footnote{Available at: \url{http://araucaria.computing.dundee.ac.uk/downloads/schemesets/katzav-reed.scm} and \url{http://assistentearg.herokuapp.com/en/schemes}.} and realize its task by believing that this constitutes a good exercise to establish and comprehend the structures and the cares one must have with each argument.

In ARG, arguments are not treated as the instantiation of an abstract inference scheme, which would refer only to the evaluation of its formal aspects. We want to provide arguments schemes and critical issues that allow the user to make stronger arguments, resistant to criticism, and avoid the incidence of reasoning faults - technical ``argument-scheme approach'' by Prakken \cite{Prakken2005}.

With the purpose of testing ARG, a pilot study was carried out in November 2014. The task was applied to 206 undergraduate law students of the Academic Center of the Rio Grande do Norte State University (UNI-RN) and involved an adaptation of the quoted case of ``the Speluncean Explorers''. The students had to give an argument, from the point of view of J. Keen, in favor of or against the maintenance of the condemnation of the accused.

After this experiment, we compared the arguments before and after the use of the ARG-system. The academics who used the tool made arguments remarkably more concise, precise, minimally structured and more resistant to criticism. In addition, the system has not received negative reviews by users, as they did not report the occurrence of error in the application. However, we need more tests for more accurate conclusions.

In sum, the advantages of ARG include: (i) being available in a scale that goes beyond the classroom; (ii)  being a useful tool for dissemination of knowledge, exposing the argumentation process in a concise way; and (iii) being a complement to face-to-face learning, as well as having online tasks, as a ``blended learning'' strategy.

\section{Final Remarks}

\begin{flushright}
\begin{minipage}{.6\textwidth}
\scriptsize
\emph{``The best we can do, that is, is to consider when we live, where we live, how we live, what the most reflective and best-informed experience of our colleagues in different areas has been, and what options are open in the future for the people who will come after us, and will revise and move beyond our ideas.  '' }\\
\t{--- Stephen Toulmin. Reasoning in Theory and Practice, 2006.}\\
\end{minipage}
\end{flushright} 

\noi ARG presents a system that can be used in any area of knowledge as an instrument of teaching critical thinking and argument production.

It is good to note that all the functionalities already implemented in ARG : (i) storing of arguments; (ii) visualization of arguments in a clear and concise way; (iii) interaction among users by tracking arguments of selected users; (iv) assistance in compounding arguments and help in improving them; (v) development of the intellectual faculties in argumentation theories; (vi) availability of multilingual scheme structures of arguments; and (vii) an evaluation system of  arguments in two dimensions (community and moderators). 

The system is available in two languages: English and Brazilian Portuguese, and is currently being translated into Spanish and French. In the future we aim to: 

\begin{itemize} 
\item Adapt anti-plagiarism solutions with the processing of language to identify similarities in the argumentative texts.
 \item Automate an exportation and importation of arguments from the professor/moderator.
 \item Integrate with a virtual environment of learning, e.g. the Moodle.
\end{itemize}

With such modifications, we hope to improve and transform it to a strong instrument for the learning of argumentation theory.

\subparagraph*{Acknowledgements}

The authors acknowledge all undergraduate law students who have contributed to the project during the Logic and Techniques of Argumentation course at UNI-RN. We would also like to acknowledge PIBIC/CNPq for their support of this research.





\bibliography{ref}

\end{document}